\documentclass[prb,twocolumn,aps]{revtex4} 
\usepackage{longtable,setspace}
\usepackage{graphicx,rotating,lscape} 
\usepackage{color}
\usepackage{float,epsfig,amsmath,amssymb,euscript,color}
\usepackage{amsmath,amssymb,epsfig,capt-of,ifthen,calc}
\usepackage{float,euscript}
\usepackage{aecompl}
\usepackage{color}
\usepackage{soul}
\usepackage{multirow}
\usepackage{dcolumn}
\usepackage{longtable}
\usepackage{supertabular}
\usepackage{setspace}
\usepackage{ulem,enumitem}
\usepackage{bm}
\usepackage{hyperref}
\usepackage{natbib}

\definecolor{bleuclair}{rgb}{0.7, 0.7, 1.0}
\definecolor{rosepale}{rgb}{1.0, 0.7, 1.0}

\begin{document}

\title{A local order metric for condensed phase environments} 

\author{Fausto Martelli$^{1}$,
\footnote[1]{Corresponding author: faustom@princeton.edu}
Hsin-Yu Ko$^{1}$, Erdal C. O\u guz$^{1,2}$ and Roberto Car$^{1,3,4,5}$}
\affiliation{%
$^{1}$Department of Chemistry, Princeton University, Princeton New Jersey, 08544 USA \\
$^{2}$Institut f$\ddot{u}$r Theoretische Physik II, Heinrich-Heine-Universit$\ddot{a}$t, D-40225 D$\ddot{u}$sseldorf, Germany \\
$^{3}$Department of Physics, Princeton University, Princeton New Jersey, 08544 USA \\
$^{4}$Princeton Institute for the Science and Technology of Materials, Princeton University, Princeton New Jersey, 08544 USA \\
$^{5}$Program in Applied and Computational Mathematics, Princeton University, Princeton New Jersey, 08544 USA \\
}%

\begin{abstract}
We introduce a local order metric (LOM) that measures the degree of order in the neighborhood of an atomic or molecular site 
in a condensed medium. The LOM maximizes the overlap between the spatial distribution of sites belonging to 
that neighborhood and the corresponding distribution in a suitable reference system. The LOM
takes a value tending to zero for completely disordered environments and tending to one 
for environments that match perfectly the reference. The site averaged LOM and its standard deviation define
two scalar order parameters, $S$ and $\delta S$, that characterize with excellent resolution crystals, liquids, and amorphous materials.
We show with molecular dynamics simulations that $S$, $\delta S$ and the LOM provide very insightful information in the study of structural transformations, 
such as those occurring when ice spontaneously nucleates from supercooled water or when     
a supercooled water sample becomes amorphous upon progressive cooling.
\end{abstract}

\keywords{}
\maketitle

\section{Introduction}
There is great interest in understanding the atomic scale transformations in processes
like crystallization, melting, amorphisation and crystal phase transitions. These processes
occur via concerted motions of the atoms, which are accessible, in principle, from molecular
dynamics simulations but are often difficult to visualize in view of their complexity. To
gain physical insight in these situations, it is common practice to map the many-body
transformations onto some space of reduced dimensionality by means of functions of the
atomic coordinates called order parameters (OPs), which measure the degree of order in a material.

Widely used OPs are the bond order parameters $Q_l$~\cite{steinhardt_bond-orientational_1983}
that measure the global orientational order of a multiatomic system from the sample average of the
spherical harmonics associated to the bond directions $\hat{\bf{r}}$ between neighboring atomic sites,
typically the nearest neighbors. The average spherical harmonics 
$\bar{Y}_{lm}\equiv \left<Y_{lm}\left(\hat{\bf{r}}\right)\right>$ depend on the choice of the reference frame 
but the bond order parameters $Q_l\equiv \left[\frac{4\pi}{2l+1}\sum_{m=-l}^{l}\left|\bar{Y}_{lm}\right|^{2}\right]^{1/2}$ 
are rotationally invariant and encode an intrinsic property of the medium. The non-zero integer $l$ specifies 
the angular resolution of the bond order parameter, with $l=6$ being a standard choice.
The $Q_l$'s take characteristic non zero
values for crystalline structures and distinguish unambiguously crystals from liquids and
glasses. In fact the $Q_l$'s vanish in the thermodynamic limit for all liquids and glasses, i.e.
systems that lack long-range order and are macroscopically isotropic. Liquids and
glasses, however, can differ among themselves in the short- and/or intermediate-range
order. Some substances, like e.g. water, exhibit polyamorphism, which means that they
can exist in different amorphous forms depending on the preparation protocol. In these
cases we would need either a measure of the local order or a measure of the global order
that could recognize different liquids and glasses. A good measure of the local order is
also crucial to analyze heterogeneous systems, such as e.g. when different phases coexist
in a nucleation process. Specializing the definition of the bond order parameters $Q_l$ to the
local environment of a site $j$ is straightforward: it simply involves restricting the
calculation of the average spherical harmonics to the environment of $j$, obtaining in this
way local bond order parameters called $q_{l}(j)$~\cite{tenWolde_numerical_1995}. However,
the $q_{l}(j)$'s are not as useful as their global counterparts, the $Q_l$'s, because they have
limited resolution and liquid environments often possess a high degree of local order that
make them rather similar to disordered crystalline environments~\cite{mickel_shortcomings_2013}. Several
approaches have been devised to improve the resolution of the measures of the local
order~\cite{tenWolde_numerical_1995,tenWolde_numerical_1996,tenWolde_homogeneous_1999,lechner_accurate_2008,ghiringhelli_state_2008,li_homogeneous_2011,kesserling_finite_2013}.
For example, it has been suggested using combinations of two or more local 
OPs~\cite{lechner_accurate_2008,volkov_molecular_2002,moroni_interplay_2005,desgranges_crystallization_2008,li_homogeneous_2011,russo_new_2014,sanz_homogeneous_2013},
but these approaches may still have difficulties in distinguishing crystalline polymorphs such as e.g. cubic (I$c$)
and hexagonal (I$h$) ices. In these situations additional analyses may be needed or one may
resort to especially tailored OPs~\cite{steinhardt_icosahedral_1981}. Examples of the latter in the context of tetrahedral
network forming systems are the local structure index (LSI)~\cite{shiratani_growth_1996,shiratani_molecular_1998}, 
the tetrahedral order parameter $q_{th}$~\cite{uttomark_kineticks_1993,chau_a_new_1998,errington_relationship_2001}, 
and the interstitial site population~\cite{russo_understanding_2013}.
\par
Recently, an alternative approach to measure the local order in materials has been
discussed in the literature, inspired by computer science algorithms known as $"$shape
matching$"$~\cite{keys_characterizing_2011}. In these schemes the similarity between an environment and a reference is
gauged by a similarity kernel, or similarity matrix~\cite{iacovella_icosahedral_2007,keys_complex_2011},
which is often represented in terms of spherical
harmonic expansions that measure angular correlations, in a way independent of the
reference frame. General theories of the
similarity kernel in the context of structure classification in materials science have been
presented in Refs~\cite{keys_complex_2011,bartok_representing_2013}. 
Approaches based on similarity kernels have been applied successfully to a number of
problems, including studies of the icosahedral order in polymer-tethered nanospheres~\cite{iacovella_icosahedral_2007},
studies of the morphology of nanoparticles~\cite{keys_complex_2011}, 
models of self-assembly~\cite{keys_characterizing_2011},
studies of quasi-crystalline and crystalline phases of densely packed tetrahedra~\cite{haji-akbari_disordered_2009},
and the prediction of the atomization energies of small organic molecules~\cite{de_comparing_2016,bartok_representing_2013}.\par
The approach that we introduce here belongs to this general class of methods and is
based on a similarity kernel of the Gaussian type~\cite{bartok_representing_2013} to measure the overlap
between a local structure and an ideal reference. In our scheme, the similarity kernel is
not represented in terms of basis functions like the spherical harmonics, but is globally
maximized by rotating the local reference after finding an optimal correspondence
between the site indices of the environment and those of the reference. Specifically, we
consider the configurations, i.e. the site coordinates, of a system of $N$ identical atoms.
The $M$ neighbors of each site define a set of local \textit{patterns}. The $M$ corresponding sites of
an ideal crystal lattice constitute the local \textit{reference}. Typically, we take $M$ equal to the
number of the first and/or the second neighbors at each site. Each pattern defines not only
a set of directions, but also a set of inter-site distances. Under equilibrium conditions the
average nearest neighbor distance takes the same value, $d$, throughout the sample and we
set the nearest neighbor distance in the reference equal to $d$. The maximal overlap of
pattern and reference at each site $j$ constitutes our local order metric (LOM) $S(j)$.\\ 
Since the overlap is maximized with respect to both rotations of the reference and permutations of
the site indices, the LOM is an intrinsic property of each local environment and is
independent of the reference frame. The LOM approaches its minimum value of zero for
completely disordered environments and approaches its maximum value of one for
environments that match perfectly the reference. The LOM is an accurate measure of the
local order at each site. It allows us to grade the local environments on a scale of ascending order defined by the
maximal overlap of each environment with the reference. In terms of the LOM we define two novel global OPs: the average 
score $S$, i.e. the site averaged LOM, and its standard
deviation $\delta S$. $S$ and $\delta S$ are scalar OPs that characterize ordered and disordered phases
with excellent resolving power.\par
In the following, we give a quantitative definition of the LOM and report an algorithm
for calculating it. We demonstrate that this algorithm maximizes the overlap between
pattern and reference in a number of important test cases. Then, we illustrate how $S$ and
$\delta S$ can be used to characterize solid and liquid phases of prototypical two- and three-dimensional
Yukawa systems, and of three-dimensional Lennard-Jones systems. Next,
we consider some more complex applications. In one of them we monitor the structural
fluctuations of supercooled water at different thermodynamic conditions within the ST2
model for the intermolecular interactions~\cite{stillinger_improved_1974}. In another we
report a molecular dynamics study of the spontaneous crystallization of supercooled
water adopting the mW model potential for the intermolecular interactions~\cite{molinero_water_2008}, showing that 
the LOM and the two global OPs $S$ and $\delta S$ provide a more accurate
description of the nucleation process than standard OPs. Finally, we report a
molecular dynamics study of the glass transition in supercooled water within the TIP4P/2005
model for the intermolecular interactions~\cite{abascal_general_2005}. This study shows that the new OPs can detect the
environmental signatures of the freezing of the translational and of the rotational motions
of the molecules.\par
The paper is organized as follows. In section~\ref{methods} we present the method.
Section~\ref{results_1} reports application to solid-liquid phase transition. 
In Section~\ref{Discerning} we apply the method to characterize the local order in water phases.
Crystallization and amorphisation of supercooled water are discussed in Section~\ref{results_2}.
Our conclusions and final remarks are presented in Section~\ref{conclusions}.
\section{Method}\label{methods}
The local environment of an atomic site $j$ in a snapshot of a molecular dynamics or
Monte Carlo simulation defines a local pattern formed by $M$ neighboring sites. Typically
these include the first and/or the second neighbors of the site $j$. There are $N$ local patterns,
one for each atomic site $j$ in the system. Indicating by $\mathbf{P}_{i}^{j} (i=1,M)$ the position vectors
in the laboratory frame of the M neighbors of site $j$, their centroid is given by
$\mathbf{P}_{c}^{j}\equiv\frac{1}{M}\sum_{i=1}^{M}\mathbf{P}_{i}^{j}$. 
In the following we refer the positions of the sites of the pattern to their centroid, i.e.
$\mathbf{P}_{i}^{j}-\mathbf{P}_{c}^{j}\rightarrow \mathbf{P}_{i}^{j}$. The local reference is the set of the same $M$ neighboring sites
in an ideal lattice of choice, the spatial scale of which is fixed by setting its nearest
neighbor distance equal to $d$, the average equilibrium value in the system of interest. For
each atomic site $j$ the centroid of the reference is set to coincide with the centroid of the
pattern, but otherwise the reference's orientation is arbitrary. Thus the positions vectors of
the reference sites in the $j$ neighborhood relative to their centroid in the laboratory frame
are given by $\mathbf{A}^{j}\mathbf{R}_{i}^{j}$ where $\mathbf{A}^{j}$ is an arbitrary rotation matrix about the centroid. The
rotation matrix can be conveniently expressed in terms of three Euler angles $\theta, \phi$ and $\psi$
belonging to the domain $\Omega\equiv \left(0\leq\theta<\pi, 0\leq\phi<2\pi, 0\leq\psi<2\pi \right)$.
The sites of the pattern
and of the reference are labeled by the indices $i$ of the position vectors. While the indices
of the reference sites are fixed, any permutation of the indices of the pattern sites is
allowed. We denote by $i_{\mathcal{P}}$ the permuted indices of the pattern sites corresponding to the
permutation $\mathcal{P}$ (if $\mathcal{P}$ is the identical permutation the pattern indices coincide with those
of the reference). For a given orientation of the reference and a given permutation of the
pattern indices we define the overlap $\mathcal{O}(j)$ between pattern and reference in the $j$
neighborhood by:
\begin{equation}
  \mathcal{O}(j)\left [\theta,\phi,\psi;\mathcal{P}\right ]=\prod_{i=1}^{M}\exp\left(-\frac{\left| \mathbf{P}_{i_{\mathcal{P}}}-\mathbf{A}^{j}\mathbf{R}_{i}^{j}\right|^2}{2\sigma^{2}M}\right)
  \label{eq:Eq1}
\end{equation}
Here $\sigma$ is a parameter that controls the spread of the Gaussian functions. Intuitively, $\sigma$ 
should be of the order of, but smaller than, $d$ for the overlap function to be able of
recognizing different environments. In our applications we adopted the choice $\sigma=d/4$ ,
as we found, in several test cases, that the results are essentially independent of $\sigma$ when
this belongs to the interval $d/4 \leq\sigma\leq d/2$ .
The LOM $S(j)$ at site $j$ is the maximum of the overlap function $\mathcal{O}(j)$
with respect to the orientation of the reference and the permutation of the pattern
indices, i.e.:
\begin{equation}
  S(j)=\max_{\theta,\phi,\psi;\mathcal{P}}\mathcal{O}(j)\left [\theta,\phi,\psi;\mathcal{P}\right ]
  \label{eq:Eq2}
\end{equation}
The LOM is an intrinsic property of the local environment at variance with the overlap
function $\mathcal{O}(j)$ that depends on the orientation of the reference and on the ordering of the
sites in the pattern. The LOM satisfies the inequalities $0 \leq S(j) \leq 1$. The two limits
correspond, respectively, to a completely disordered local pattern ($S(j)\rightarrow 0$) and to an
ordered local pattern matching perfectly the reference ($S(j)\rightarrow 1$). The LOM grades each
local environment on an increasing scale of local order from zero to one. 
As a consequence of the point symmetry of the reference the overlap function defined in Eq.~\ref{eq:Eq1}
has multiple equivalent maxima. We present in Sect~\ref{Optimization} an effective optimization
algorithm to compute $S(j)$.
We define two global order parameters based on $S(j)$. One is the average score $S$ or site
averaged LOM:
\begin{equation}
  S=\frac{1}{N}\sum_{j=1}^{N}S(j)
  \label{eq:Eq3}
\end{equation}
The other is the standard deviation of the score that we indicate by $\delta S$ :
\begin{equation}
  \delta S = \sqrt{\frac{1}{N}\sum_{j=1}^{N}\left(S(j)-S\right)^{2}}
  \label{eq:Eq4}
\end{equation}
In the following sections of the paper we show with numerical examples that the score $S$
has excellent resolution and is capable of characterizing with good accuracy the global 
order of both crystalline and liquid/amorphous samples. The standard deviation of
the score, $\delta S$, provides useful complementary information and can enhance the
sensitivity of the measure of the global order in the context of structural transformations.
\subsection{Optimization algorithm}\label{Optimization}
The overlap function $\mathcal{O}(j)$ defined in Eq.~\ref{eq:Eq1} has $L$ equivalent maxima. Here $L$ is the
number of proper point symmetry operations of the reference. If a maximum corresponds
to the permutation $\bar{\mathcal{P}}$ of the pattern indices and to the Euler angles $(\bar{\theta},\bar{\phi},\bar{\psi})$, 
all the other distinct but equivalent maxima can be obtained from the known maximum by rotating the
reference from the direction $(\bar{\theta},\bar{\phi},\bar{\psi})$ with the $L-1$ point symmetry operations different
from the identity, and by updating correspondingly the permutations of the pattern
indices. To compute $S(j)$ (Eq.~\ref{eq:Eq2}) it is sufficient to locate only one of these maxima. In
view of the point symmetry of the reference, it would be sufficient for that to explore
only a fraction $1/L$ of the Euler angle domain $\Omega$, which we may call $\Omega/L$ , the
irreducible domain of the Euler angles. In the present implementation, however, we opted to explore the full 
$\Omega$ domain for reasons of simplicity. We also notice that $\mathcal{O}(j)$ in Eq.~\ref{eq:Eq1} decays rapidly to
zero when the distance between any one of the pattern sites and the corresponding
reference site is sufficiently larger than $\sigma$. 

To identify a permutation for which $\mathcal{O}(j)$ is in the neighborood of a maximum
we proceed as follows. First, we represent the domain $\Omega$
on a uniform grid with $K/2 \times K \times K$ points. Here $K$ is an even integer equal to the
number of intervals dividing the range of each Euler angle. The grid points $(k_1, k_2 , k_3)$
with $(k_1=0,K/2-1, k_2=0,K-1, k_3=0,K-1)$, define the set of Euler angles
$(\theta=2k_1\pi/K, \phi=2k_2\pi/K, \psi=2k_3\pi/K)$, the corresponding rotation matrices $\mathbf{A}^{j}$ and the
associated reference vectors $\mathbf{A}^{j}\mathbf{R}_{i}^{j}$
that we shall use in Eq.~\ref{eq:Eq1}. Next, we select an initial
grid point, such as e.g. $(k_1=k_2=k_3=0)$. For this grid point we select a permutation $\mathcal{P}$
with the following algorithm: 1. Pick one of the $M$ pattern sites at random with uniform
probability: let it be site $i'$. 2. Compute the square distances $\left|\mathbf{P}_{i'}^{j}-\mathbf{A}^{j}\mathbf{R}_{i}^{j}\right|^{2}$
between the pattern site $i'$ and all the $M$ reference sites. 3. Set $i'$ equal to the index of the reference
site with the smallest distance from $P_{i'}^{j}$.
4. Pick at random with uniform probability one
of the remaining $M-1$ pattern sites and compare it with the remaining $M-1$ reference
sites. 5. Set the index of the chosen pattern site equal to that of the reference site with the
smallest distance from it. 6. Pick at random one of the remaining $M-2$ sites and
compare it with the remaining reference sites in order to set its index. 7. Proceed in the
same way until all the indices have been set. Having selected a permutation $\mathcal{P}$ we compute $\mathcal{O}(j)$ with 
Eq.~\ref{eq:Eq1}. Then we move to a neighboring point on the grid of the Euler angles, e.g. $(1,0,0)$,
repeat the above procedure to select a permutation of the indices starting from the
permutation selected at the previous step, and compute the corresponding $\mathcal{O}(j)$. We repeat
this procedure by moving systematically on the grid of the Euler angles (e.g. sample $k_1$
first, followed by $k_2$, and then by $k_3$). The largest $\mathcal{O}(j)$
calculated in this way is guaranteed to be in the basin of attraction of one of the
equivalent maxima of $\mathcal{O}(j)$ and the subsequent optimization can be done with respect to
the Euler angles only, while keeping the permutation of the indices fixed. This angular
optimization is a local optimization problem and can be performed effectively with
gradient methods, such as steepest descent or conjugate gradients.

In all our applications we adopted the choice $K=18$, which we found to be sufficient
for good convergence. To check that the algorithm leads correctly to the maximum of $\mathcal{O}(j)$
we made several tests. In some of them we considered a perfect crystalline environment
(at zero temperature) and chose a reference based on the same crystalline structure. In
this case $S(j)$ should take the value $S(j)=1$. We found that this was always the case when
starting from random permutations of the pattern sites and random orientations of the
reference. In other tests we considered disordered crystalline and liquid environments at
different temperatures. In these cases the exact values of $S(j)$ are not known a priori.
However, in all the cases we found that $S(j)$ converged always to the same value within the tolerance 
of the convergence criterion, independently of the initial random values chosen for the permutation of 
the pattern indices and for the orientation of the reference.

\begin{figure}[!]
  \begin{center}
   \includegraphics[scale=.32]{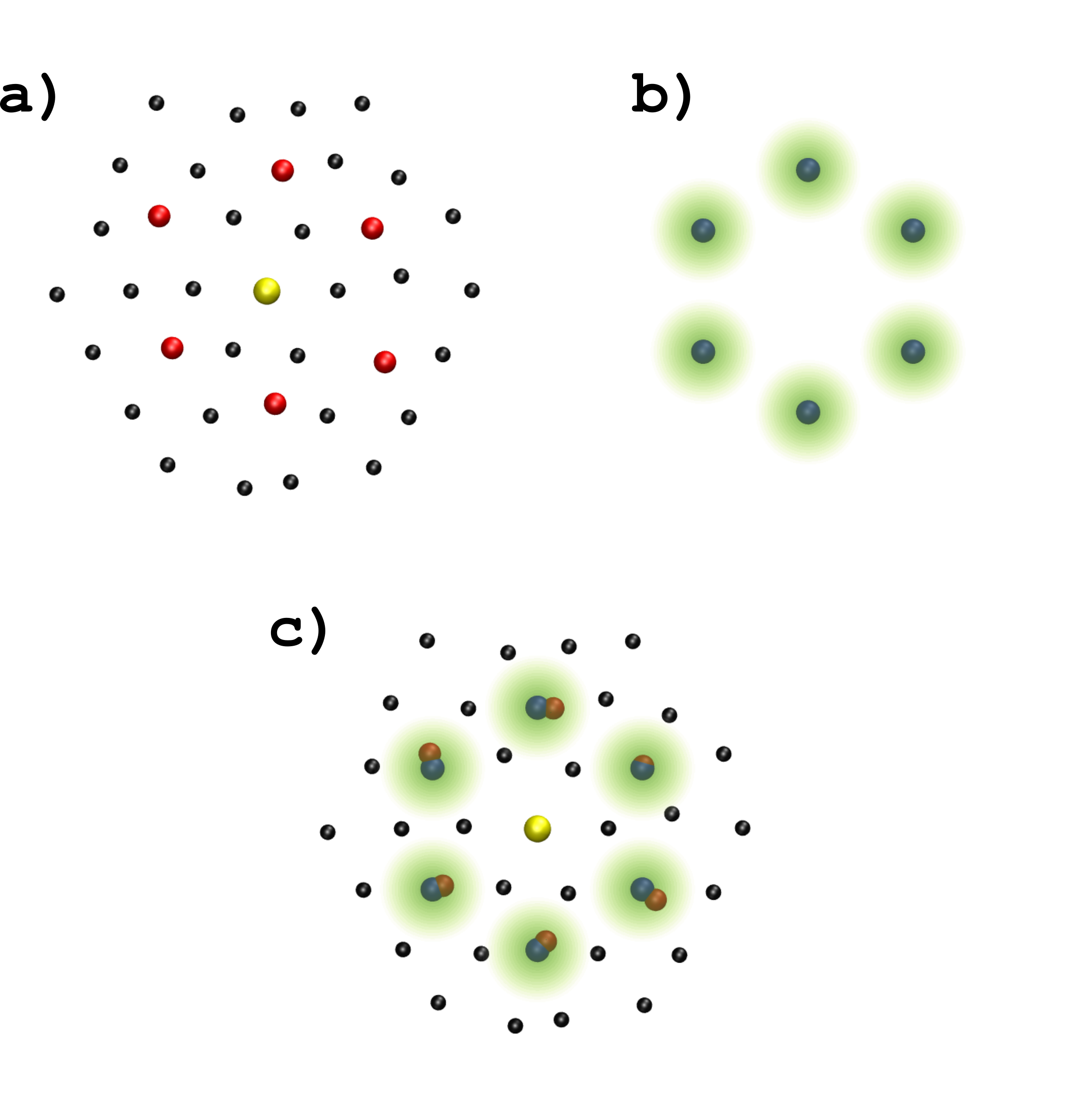}
   \caption{Schematic representation of our approach in an application
   to a 2D Yukawa crystal at $k_{B}T/V_0=0.001$. a):
   The red spheres indicate the second shell of neighbors of site
   $j$ (yellow sphere). b): The $6$ blue spheres are the vertices of
   the reference hexagon. The green shaded areas represent the
   Gaussian domains. c): Optimized overlay of reference and
   local patterns.}
   \label{fig:Fig1}
  \end{center}
\end{figure}

In Fig.~\ref{fig:Fig1} a two-dimensional crystal with Yukawa pair interactions is used to illustrate
the method. The system has been equilibrated at finite temperature. As reference we
choose the $6$ sites associated to the second shell of neighbors in the ideal triangular
lattice. The $6$ sites are the vertices of a regular hexagon. The picture shows (a) a local
environment, (b) the corresponding reference with shaded areas representing the regions
in the neighborhood of the reference selected by $\sigma$, (c) the optimal overlap between
pattern and reference for the local environment depicted in (a).

\section{Applications to simple systems}\label{results_1}
As a first application we use the new OPs to analyze simulations of simple condensed phase systems at varying temperature.
Initially the temperature is low and the systems are in the solid state. 
When the temperature exceeds a certain threshold, the solid looses mechanical stability and the atomic dynamics 
becomes diffusive signaling a transition to the liquid state. We have considered, in particular, the following systems: 
a 2D system of identical particles with Yukawa pair interactions, a 3D system of identical particles with Yukawa pair
interactions, and a 3D system of identical particles with Lennard-Jones pair interactions. In all cases we find 
that the new OPs signal the transition to the liquid state with sensitivity equivalent to that of popular OPs, 
like $Q_6$ and its two-dimensional specialization $\Psi_6$~\cite{chakrabarti_effect_1998}. 
However, once in the liquid state our OPs can still quantify the degree of order 
and are therefore superior to $Q_6$ and $\Psi_6$. 
\subsection{Yukawa system in 2D}\label{yuk2d}
Here we perform Brownian dynamics simulations 
of particles with repulsive pair interactions given by the Yukawa potential $V(r)=V_0 \exp(-\kappa r) / \kappa r$, 
where $r$ denotes the inter-particle separation, and $\kappa$ is the inverse screening length. The strength of the interaction 
is set by the amplitude $V_0$. We consider a system of $9180$ particles in the $NVT$ ensemble
with periodic boundary conditions. We start the simulations from a perfect triangular lattice.
We analyse the degree of local order as a function of $k_BT/V_0$ at the fixed reduced screening length
$\rho/\kappa^2 = 0.21$, where $\rho$ is the 2D number density. Without loss of generality we hereby 
use $V_0 = 1$. Pattern sites comprise the second shell of neighbors and are compared with the reference 
in Fig.~\ref{fig:Fig1} for a representative snapshot of the solid at $k_BT/V_0=0.001$. 
\begin{figure}[!]
  \begin{center}
   \includegraphics[scale=.32]{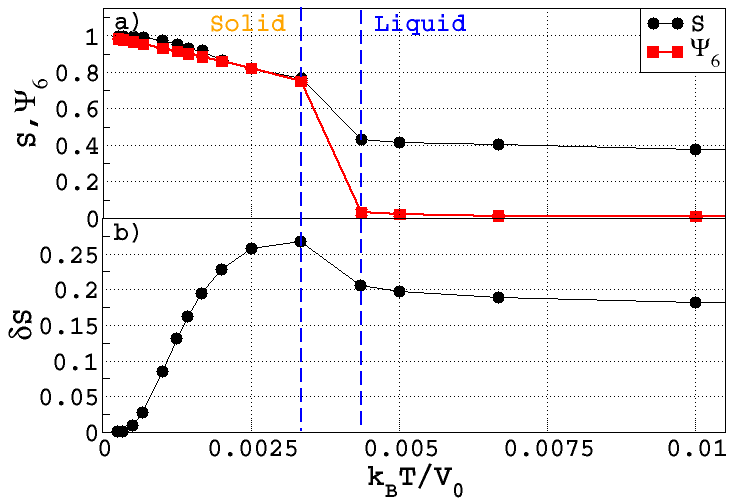}
   \caption{ a): 2D Yukawa system. Profile of $S$ (black dots) and of $\Psi_{6}$ (red squares) as a function 
            of $k_BT/V_0$. The blue dashed lines delimit the region of instability of the solid phase in the simulation.
            b): 2D Yukawa system. Profile of $\delta S$ as a function of $k_{B}T/V_0$.} 
   \label{fig:Y2}
  \end{center}
\end{figure}
Panel (a) of Fig.~\ref{fig:Y2} compares the global OPs $S$ (black dots) and $\Psi_{6}$ (red squares). 
As expected, both $S$ and $\Psi_6$ for the perfect crystal take the value of $1$. 
As $T$ increases, both $S$ and $\Psi_6$ decrease.
In correspondence with the blue dashed lines, signaling instability of the crystal, both OPs show a quick drop. 
In the liquid phase,
$\Psi_6\simeq 0$, as expected, while 
$S$ keeps a finite value of $S\simeq0.4$, which slightly decreases as the temperature is further increased.
Therefore, both $S$ and $\Psi_6$ identify the phase transition, but only $S$ is able to quantify the degree of 
order remaining in the liquid phase.
 
The behavior of $\delta S$ in panel (b) gives further insight on the solid-liquid transition.
$\delta S$ takes the maximum value in the crystal at the highest temperature and drops substantially
in the liquid phase. This behaviour follows from
the non linear nature of the LOM. 
The liquid phase has more strongly disordered local patterns with sites often belonging to the tails of the Gaussian
domains in Eq.~\ref{eq:Eq1}. Site fluctuations in the liquid weight less than fluctuations in the solid,
where patterns sites are closer to the centers of the Gaussian domains. 

We have observed the same behavior of $\delta S$ in all the 
solid-liquid transitions that we have investigated, namely $\delta S$ takes its maximum value in the hot crystal 
before the occurrence of the dynamical instability that signals melting. It is tempting to notice the similarity 
of this behavior with Lindemann's melting criterion~\cite{lindemann_calculation_1910}, according to which 
melting occurs when the average atomic displacement exceeds some fraction of the interatomic distance. In our 
approach the dynamic instability is associated to the largest fluctuation of $S$.\\
\subsection{Yukawa and Lennard-Jones systems in 3D}\label{crys}
\begin{figure}[!]
 \centering
    \includegraphics[scale=.32]{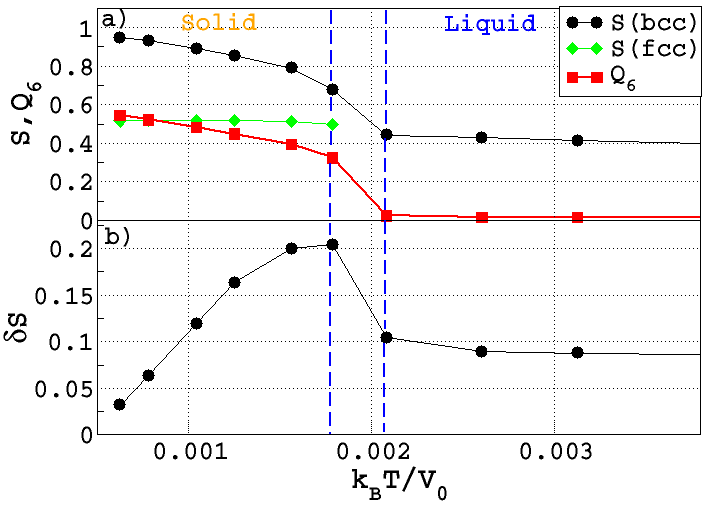}
    \vspace{-0.40cm}
    \includegraphics[scale=.32]{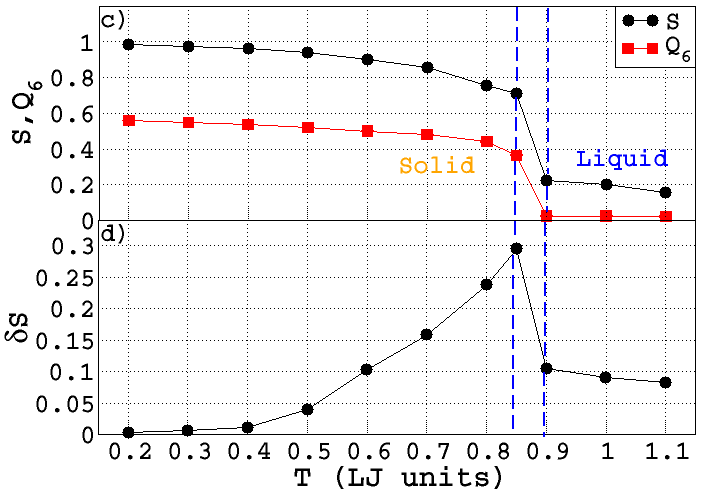}
    \caption{ a): $S$ computed with ideal bcc crystal reference (black dots), $Q_6$ (red squares), and $S$ computed with 
          fcc reference (green diamonds), as a function of $k_BT/V_0$ for a 3D Yukawa system. 
          b): $\delta S$ corresponding to $S$ with bcc reference for the same Yukawa system. 
          c): $S$ (black dots) and $Q_6$ (red squares) as a function of $T$ for a Lennard-Jones system.
          d): $\delta S$ for the same Lennard-Jones system.
          The vertical blue dashed lines delimit the regions of crystal instability in the simulations.}
 \label{fig:Y3}
\end{figure}

In Fig. \ref{fig:Y3} we report
$S$ and $Q_6$ as a function of the temperature for a 3D system of identical particles with Yukawa pair interactions 
(panels (a) and (b)) and for a 3D system of identical particles with Lennard-Jones pair interactions 
(panels (c) and (d)). At low temperature the Yukawa system is in the bcc crystalline phase whereas the Lennard-Jones 
system is in the fcc crystalline phase. 
In the Yukawa system we use the pairwise interactions introduced in Section~\ref{yuk2d}.
We sample the $NVT$ ensemble with Brownian dynamics.
The simulation cell contains $4394$ particles with periodic boundary conditions. 
The reference includes the
first and the second shell of neighbors of a perfect bcc lattice for a total of $14$ sites. 
Panel (a) shows $S$ (black dots) and $Q_6$ (red squares) versus temperature. 
At very low temperatures, $S\simeq 1$ because reference and pattern overlap almost perfectly. Increasing
the temperature, both OPs show a quick drop in correspondence with the phase transition 
with $Q_6\rightarrow 0$ as expected in the liquid phase, and $S$ taking a value $S\simeq 0.4$. 
Like in 2D, both order parameters are able to identify the phase transition, 
but $S$ provides quantitative information on the order present in the liquid, 
whereas $Q_6$ takes zero value in all liquids in the thermodynamic limit.
It is worth recalling that local OPs, such as the $q_l(j)$, have difficulties in identifying the bcc symmetry of 
hot crystals before melting~\cite{da-qi_structure_2005}.
Our approach uses a non-linear LOM, which can unambiguously distinguish distorted
local bcc structures from distorted local fcc or hcp structures. 
To illustrate this statement we report in the temperature range of crystalline stability in Fig.~\ref{fig:Y3} (a) 
the profile of $S$ (green diamonds) obtained by using as reference the first shell of neighbors of the fcc lattice. 
We notice that $S$(bcc) and $S$(fcc) are well separated in the solid phase even at the highest temperatures. 
Panel (b) shows $\delta S$. As in 2D, $\delta S$ takes its maximum value in the solid phase before the onset 
of crystalline instability in the simulation.

In panels (c) and (d) we report the same data for a 3D system of identical particles interacting with the 
Lennard-Jones potential with parameters appropriate to Argon~\cite{frenkel2001understanding}.
In this case we perform Monte Carlo simulations in the $NpT$ ensemble with a periodic box containing $1372$ particles.
We choose as reference the anticuboctahedron, which has $12$ vertices, and corresponds to the first shell
of neighbors in the ideal fcc lattice. 
The temperature variation of $S$ (black circles) and $Q_6$ (red squares) is shown in panel (c).
At very low temperature, $S\simeq 1$ due to the nearly perfect overlap of patterns and reference. $S$ and $Q_6$
are able to distinguish the crystalline solid from the liquid, and show a substantial drop in correspondence with the 
dashed vertical blue lines. In the liquid, $Q_6\rightarrow0$  
as expected, while $S$ remains finite with a value close to $0.2$. 
Similarly, $\delta S$ in panel (d) takes its maximum value in the crystal at the highest temperature.

\begin{figure}[!]
  \begin{center}
   \includegraphics[scale=.15]{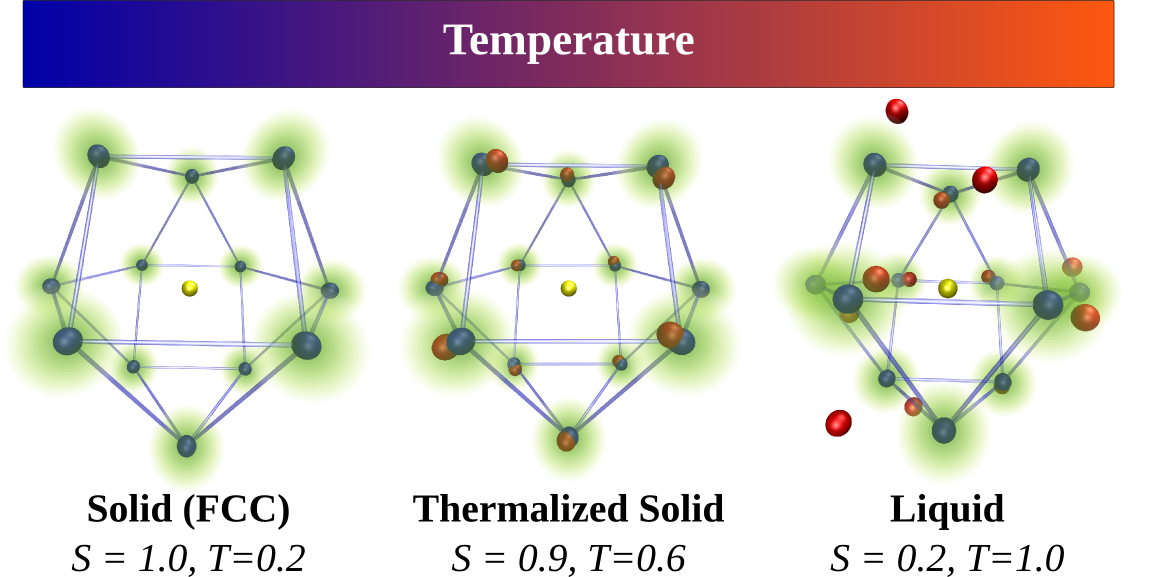}
   \caption{Local environments in solid and liquid Lennard-Jonesium at different temperatures. The notation is the same of
   Fig.~\ref{fig:Fig1}. Neighboring reference sites have been connected by thin lines to emphasize the structure of the 
   anticuboctahedron.}
   \label{fig:Ar_snap}
  \end{center}
\end{figure}
Representative local environments at different temperatures around a site indicated by a yellow sphere are shown
in Fig.~\ref{fig:Ar_snap}: 
the environment on the left corresponds to a cold crystal ($T=0.2$), the one in the middle to a hot crystal ($T=0.6$), and the one on the right to a liquid ($T=1.0$).
One may notice
the increasing deviation with temperature of the pattern sites (red spheres) relative to the reference sites
(blue spheres). In the liquid state some of the pattern sites move in the tail region of the Gaussian domains, causing
a drop in both $S$ and $\delta S$.
\section{Local structures in water phases}\label{Discerning}
Molecular systems like water exhibit a rich phase diagram, with two competitive crystalline phases,
cubic (I$c$) and hexagonal (I$h$) ice, respectively, at low pressure. Moreover, metastable undercooled liquid water 
transforms continuously with pressure from a low-density form (LDL) to a high-density one (HDL)~\cite{soper_structures_2000}.
In the following we consider representative I$c$ and I$h$ solids, LDL and HDL 
liquids at different thermodynamic conditions. 

Water molecules bind together by hydrogen bonds forming a tetrahedral network connecting 
neighboring molecules.
To describe this network it is sufficient to consider the molecules
as rigid units centered on the oxygens. The sites that define the local order are the oxygen sites and 
application of the formalism is straightforward. 
Water structures are dominated by tetrahedral hydrogen bonds and have similar short range
order (SRO). The intermediate range order (IRO) is significantly more sensitive to
structural changes than the SRO. We choose therefore references associated to the second
shell of neighbors in crystalline ices. In particular, we adopt either the cuboctahedron $(C)$
or the anticuboctahedron $(\bar{C})$, both of which have $12$ vertices and correspond to the
second shell of neighbors in cubic and hexagonal ices, respectively.
In these simulations we used the ST2 force field~\cite{stillinger_improved_1974} for water with periodic boundary 
conditions and adopted the Ewald technique (with metallic boundaries) to compute the electrostatic sums.
\subsection{Hexagonal and cubic ice}
\begin{figure}[!]
  \begin{center}
   \includegraphics[scale=.38]{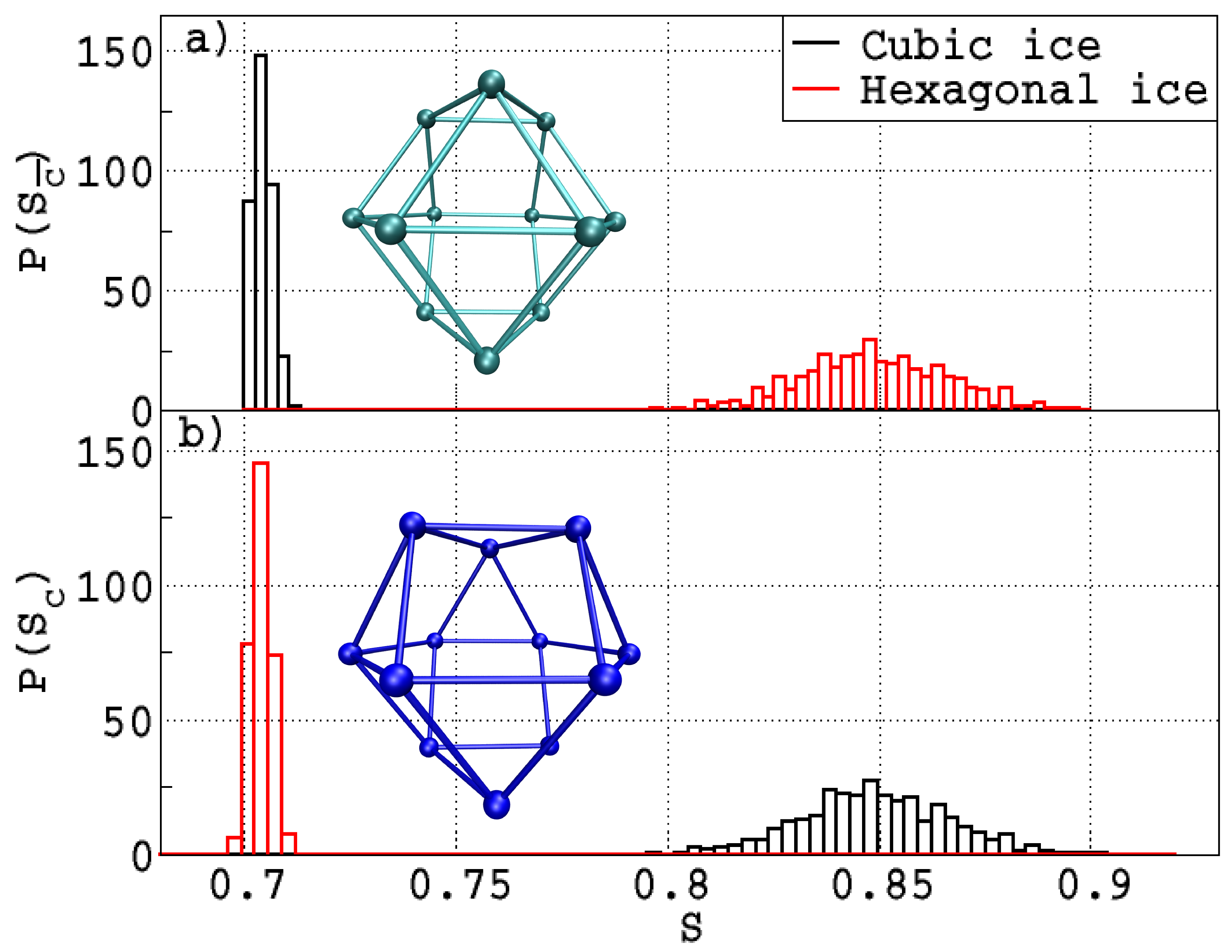}
   \caption{a): distribution of $S$ for I$c$ ice (black) and for I$h$ ice (red) with the $\bar {C}$ reference.
            b): distribution of $S$ for I$c$ ice (black) and for I$h$ ice (red) with the $C$ reference.
            The $\bar {C}$ and the $C$ references are depicted in the upper and in the lower panel, respectively. 
            Spheres representing oxygen atoms are connected by sticks to emphasize the cuboctahedron (green) and 
            the anticuboctahedron (blue).}
   \label{fig:distrib}
  \end{center}
\end{figure}

The simulation box for I$c$ ice is cubic and contains $512$ molecules, while that for I$h$ ice is orthorhombic and contains
$768$ molecules. We thermally equilibrate both solids via classical MD simulations in the $NpT$ ensemble 
at $T=250$ K and $p=1$ bar.
In panel (a) of Fig.~\ref{fig:distrib} we report the distribution
of $S$ with reference $\bar {C}$ indicated by $S_{\bar C}$ for cubic (black) and hexagonal (red) ices. 
In panel (b) of the same figure, 
we report the corresponding distributions of $S$ with reference $C$ indicated by $S_C$.
It is clear from both panels that the distributions based on the two different references for the same crystal are 
well separated. Moreover, the distributions corresponding to the two different crystals are also well separated 
irrespective of the reference we use.

One notices in Fig.~\ref{fig:distrib} that the distribution of the order parameter $S$ is broad when the reference
is based on the same lattice of the pattern, i.e. $\bar C$ for I$h$ ice and $C$ for I$c$ ice.
The distribution is instead rather sharp when the $\bar C$ reference is used to measure I$c$ patterns or
when the $C$ reference is used to measure I$h$ patterns. 
This behavior is a consequence of the non-linearity of the LOM. When pattern and reference correspond to the
same crystalline lattice the pattern sites are closer to the reference sites and small fluctuations in the pattern
cause relatively large variations of the LOM. On the other hand, when pattern and reference do not correspond to the same 
crystalline lattice, pattern sites deviate more from the reference sites and small fluctuations in the pattern cause
relatively small variations of the LOM.
\subsection{Low-density and high-density liquid water}
We performed MD simulations for water in the $NpT$ ensemble at $T=240$ K and $p=1$ bar and $p=3$ kbar, respectively.
The case $p=1$ bar is representative of a LDL liquid, while the case with $p=3$ kbar is representative of a HDL liquid.
We use a cubic box containing $512$ molecules with periodic boundary conditions.

\begin{figure}[!]
  \begin{center}
   \includegraphics[scale=.40]{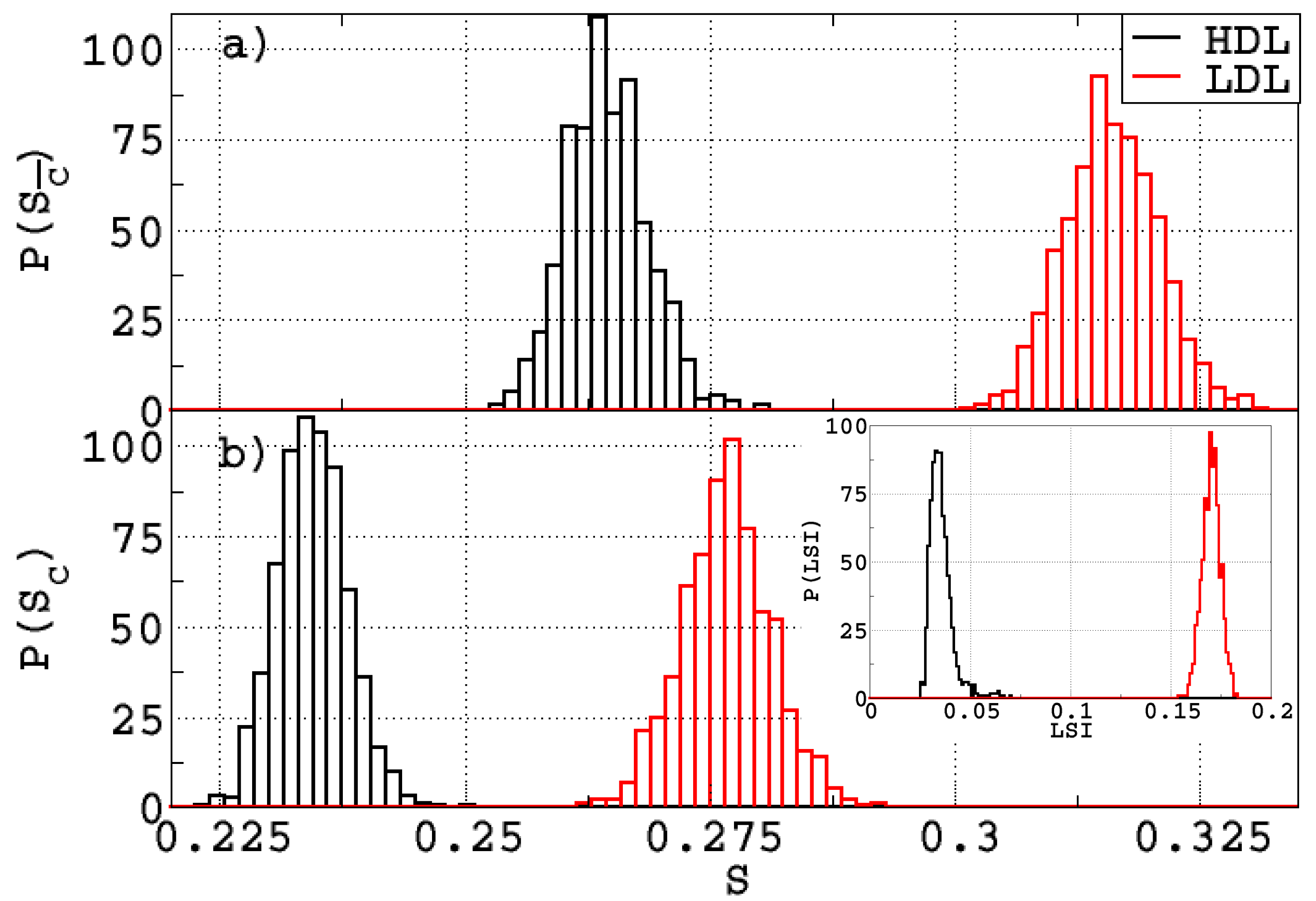}
   \caption{a): distribution of $S$ for HDL water (black) and for LDL water (red) with the $\bar {C}$ reference.
            b): distribution of $S$ for HDL water (black) and for LDL water (red) with the $C$ reference.
            The $\bar {C}$ and the $C$ references are depicted in Fig.~\ref{fig:distrib}.
            The inset shows the distribution of LSI for HDL water (black) and LDL water (red). }
   \label{fig:histo}
  \end{center}
\end{figure}

Resolving the local order in disordered structures, like HDL and LDL water, is difficult.  
Standard local OPs such as $q_6(j)$ fail in this respect and ad hoc OPs like the local structure index (LSI) have been
devised for the task. The LSI is sensitive to the order in the region between the first two
coordination shells of water. In this region the LSI detects the presence of interstitial molecules,
whose population increases as the density or the pressure increases. 
While the LSI is an OP especially tailored for water, $S$ is non specific to water but has 
resolving power equivalent to that of the LSI in liquid water, as illustrated in
Fig.~\ref{fig:histo}. The two panels in this figure show the distribution of $S_{\bar C}$ (panel (a)) and of 
$S_C$ (panel (b) for HDL (black) and LDL (red).
In both cases the distributions are well separated, similarly to the LSI distributions shown in the inset
in panel (b).
Independently of the adopted reference, $S$ has a higher value in LDL than in HDL, reflecting the higher degree 
of order in the former. 
By comparing the two panels in Fig.~\ref{fig:histo} we also see that both liquids have higher $\bar{C}-$ than $C-$character.
Both LDL and HDL structures are well distinct from the crystalline
reference and the corresponding broadening of the $S$ distributions is approximately the same in the two liquids.
\section{Crystallization and amorphisation of supercooled water}\label{results_2}
To further illustrate the power of the LOM and of $S$ and $\delta S$ 
we consider the complex
structural rearrangements occurring in supercooled water during  crystallization or when a liquid sample amorphizes 
under rapid cooling. 

To model crystallization, we consider rigid water molecules interacting with the mW
potential~\cite{molinero_water_2008}. The mW potential describes the tetrahedrality of the molecular arrangements, 
but does not have charges associated to it missing the donor/acceptor character of the hydrogen bonds. 
For that reason crystallization occurs much faster with mW than with more realistic potentials that describe more accurately
the hydrogen bonds. At deeply supercooled conditions 
mW water crystallizes spontaneously on the time scale of our molecular dynamics simulations. In spite of the simplified
intermolecular interactions in mW water, ice nucleation is a very complex process and access to good order parameters 
is essential to interpret the simulations. 

To model amorphisation we adopt the more realistic 
TIP4P/2005 potential~\cite{abascal_general_2005} for the intermolecular interactions. 
This potential applies to rigid molecules but takes into account the charges associated to the hydrogen bonds.
This level of description is important
to model the relaxation processes that occur in a liquid sample undergoing amorphisation. The processes 
that lead to the freezing of translational and rotational degrees of freedom in the glass
transition are captured well by our OPs.   

\subsection{Crystallization of supercooled water} \label{Crystallization}
\begin{figure}[!]
 \centering
    \includegraphics[scale=.33]{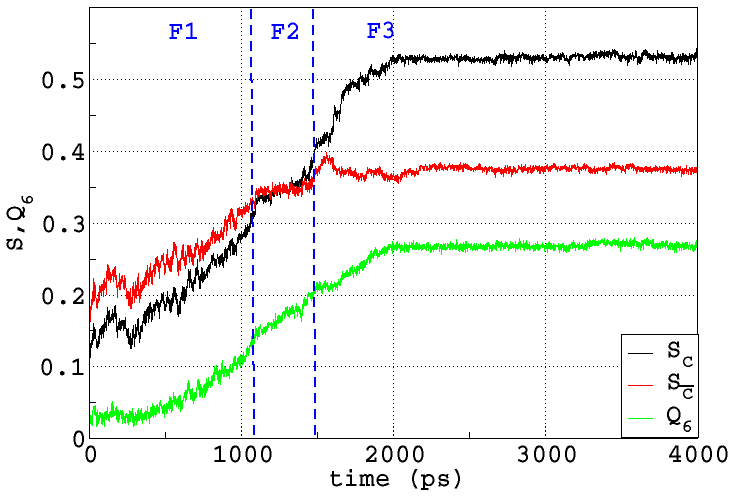}
    \caption{Comparison of $S_{C}$ (black), $S_{\bar C}$  (red), and $Q_6$ (green) 
          as a function of time in the crystallization of $1000$ water molecule interacting via the mW potential.
          The blue vertical dashed lines indicate three time frames F1, F2, and F3, respectively.}
 \label{fig:mW}
\end{figure}

To study crystallization we performed classical MD simulations in the $NVT$ ensemble, using $1000$ molecules
with interactions described by the mW potential~\cite{molinero_water_2008} in a parallelepipedic box with side lengths
ratios $L_{z}/L_{x}=4,L_{y}/L_{x}=1$, and periodic boundary 
conditions. We set the temperature to $T=190$ K and the volume of the box to a mass density 
of $\rho=0.98$g/cm$^3$. At these thermodynamic conditions
spontaneous crystal nucleation occurs rapidly in the mW fluid~\cite{molinero_water_2008}. 
The evolution of the water sample starting from an equilibrated liquid is illustrated in Fig.~\ref{fig:mW}, 
\begin{figure}[!]
 \centering
    \includegraphics[scale=.33]{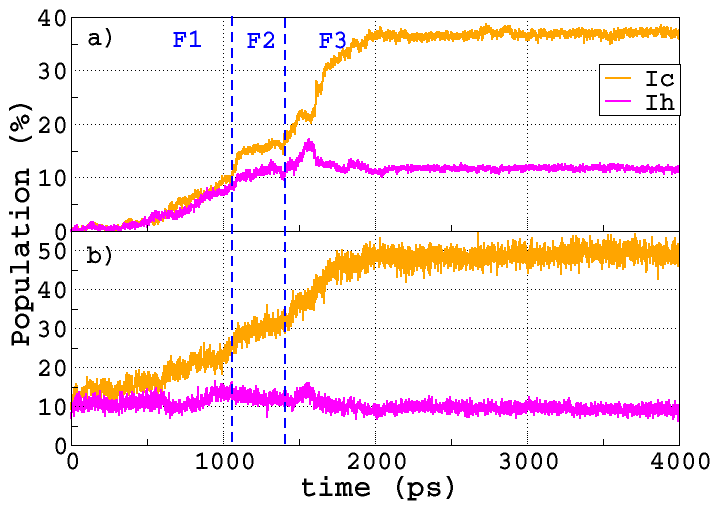}
    \caption{Time evolution during crystallization of mW water of the relative population of the sites with crystalline 
    $I_c$ character (orange line) and of the sites with crystalline $I_h$ character, as determined from the LOM 
    (a) and from $q_6/w_4$ (b). See text for a detailed explanation.}
 \label{fig:mW_2}
\end{figure}
where we report the evolution with time of three OPs: $Q_6$ (green line), $S_C$ 
(black line) and $S_{\bar C}$ (red line), respectively.
In this figure we recognize three time frames separated by the dashed vertical blue lines and indicated by 
F1, F2, and F3, respectively. In F1 the system is liquid but becomes increasingly structured 
as indicated by the growth of the three OPs. Microcystallites keep forming and disappearing. 
$S_C$ and $S_{\bar C}$ are more sensitive than $Q_6$ to the fluctuations 
of the local order, as indicated by the greater fluctuations of the black and red lines relative to the green line 
in F1. Interestingly, the liquid has stronger $\bar{C}-$ than $C-$character, in accord with
Fig.~\ref{fig:histo}. The relative weight of the $\bar C$- and $C$- characters reverses as crystallization proceeds.
F2 marks the appearance of a stable crystallite that further grows in the 
initial stage of F3. This complex kinetics is not captured by $Q_6$ (green line) which shows only a continuous growth with
time. Instead, both $S_C$ and $S_{\bar C}$ identify a plateau in F2, in correspondence with the formation of a 
stable crystallite.

The nucleating ice is a mixture of cubic and hexagonal ices, with a prevalence of the former, as indicated by the 
larger overall growth of $S_C$ in the simulation. Indeed, during the entire evolution shown in Fig.~\ref{fig:mW}, 
$S_C$ varies more than $S_{\bar C}$.
Due to $NVT$ sampling with periodic boundary conditions, liquid water is 
always present in the sample and does not disappear even when the nucleation process is completed in F3. 
The residual liquid water has more $\bar{C}-$ than $C$-character and, therefore, the $S_C$ and 
the $S_{\bar C}$ profiles should be considered merely as qualitative site-averaged contours.

More quantitative insight can be extracted from Fig.~\ref{fig:mW_2}, where we report the
time evolution of the fraction of cubic and hexagonal sites.
This analysis is based on the LOM and is independent of the reference choice since distributions of the competing 
ice and liquid structures in Figs.~\ref{fig:distrib} and ~\ref{fig:histo} do not overlap. Thus, in the 
remaining part of this section and in Sect. VB as well, we use the $C$ reference and omit from $S$ the corresponding 
subscript. We introduce two cutoff values, $S_1 = 0.6$ and $S_2 = 0.75$, to distinguish the local 
environments. If at site $j$ the LOM satisfies $S^{j}<S_1$ the local
environment is liquid-like, if $S_1< S^{j}<S_2$ the local environment is ice hexagonal-like, and 
if $S^{j} >S_2$ the local environment is ice cubic-like. Notice that the results do not depend on the actual 
values of the cutoffs $S_1$ and $S_2$ as long as they fall inside regions where the $S$ distribution has negligible weight. 
The time evolution of the fraction of sites with cubic and hexagonal character resulting from the LOM 
is reported in panel (a) of Fig.~\ref{fig:mW_2}.
In the F1 frame both cubic and hexagonal fractions grow with a slight dominance of the former.  
This growth is associated to crystallites that keep forming and disappearing. In 
correspondence with the first dashed vertical line the growth becomes faster for both environments, signaling
the formation of a stable crystalline nucleus with mixed character, in which I$c$ and I$h$ sites are 
separated by a stacking fault. At this point hexagonal growth almost entirely stops while 
cubic ice continues to grow at a slower pace by incorporating nearby crystallites with the same character.
The stable nucleus contains approximately $300$ out of $1000$ sites and takes the form of a large ice cluster embedded in 
a dominant liquid environment. Towards the end of the F2 frame, cubic ice growth accelerates 
and, in the early stage of the F3 frame, the size of the crystalline cluster rapidly reaches the size of the box. 
At this point no further growth is possible. 
In the early stage of F3 hexagonal growth is significantly less pronounced than cubic growth and is mainly associated 
to a visible hump shortly after the onset of F3.
The hump is due to small clusters with hexagonal character
that form on the surface of the large cubic crystallite, and then rapidly convert to cubic character. 
The nucleation ends with the formation of a large crystallite that spans the size of the box and includes $\sim 50\%$ 
percent of the available sites. Of the crystalline sites $\sim 80\%$ percent have cubic and 
$\sim 20\%$ percent hexagonal character.
The quantitative details of the nucleation process depend on the MD trajectory. For instance the
relative fraction of cubic and hexagonal sites changes from one trajectory to another. Qualitatively, however,
the process is the same in all the 10 trajectories that we have generated. 
Our results are quantitatively very similar to a previous analysis in which cubic and hexagonal sites were identified in 
terms of eclipsed and staggered local configurations~\cite{moore_freezing_2010,moore_is_2011,nguyen_identification_2015}.
To further illustrate the superior resolving power of the LOM relative to common OPs we report in panel (b) of 
Fig.~\ref{fig:mW_2} an analysis of the same MD trajectory of panel (a) using a combination of the two orientational OPs 
$q_6(j)$ and $w_4(j)$~\cite{steinhardt_icosahedral_1981}. In this approach, $q_6(j)$ is extracted from the nearest neighbors 
of a site $j$ and serves to determine the liquid or the crystalline character of the site. If the site $j$ is crystalline 
one assigns to it cubic or hexagonal character depending on the value
of $w_4(j)$, whose computation requires the first and second neighbors of the site $j$. There are important quantitative 
differences between panel (a) and panel (b) of Fig.~\ref{fig:mW_2}. A major difference is already apparent in the time 
frame F1: in panel (b) a significant fraction of the sites that are considered liquid in panel (a) are classified 
as crystalline sites since the very beginning of the trajectory. This is due to the fact that liquid- and crystal-like
configurations overlap in the $q_6$ distribution. Similarly, the relative fractions of cubic and hexagonal sites 
of panel (a) at the end of the trajectory is not reproduced well in panel (b), again because of the overlap 
of cubic and hexagonal configurations in the $w_4$ distribution.
\subsection{Amorphisation of supercooled water} \label{glass_transition}
To study amorphisation, we have performed classical MD simulations for a system composed of $216$ molecules 
interacting via the TIP4P/2005 potential~\cite{abascal_general_2005} in a cubic simulation box with periodic 
boundary conditions. Starting from an equilibrated
liquid at $240$ K and $p = 1$ bar, we performed isobaric cooling with a rate of $5$ K/ns 
to generate an amorphous ice structure. Given the adopted protocol this structure should have similarity 
to experimentally prepared low density amorphous (LDA) ice structures~\cite{mayer_new_1985}.
Our cooling rate is slightly higher then the one recently adopted in molecular dynamics simulations for the same 
water model~\cite{wong_pressure_2015}. However, our goal is not to generate a high quality amorphous structure, 
but rather to test whether our approach can be used to study the glass transition in water.  
\begin{figure}[!]
 \centering
    \includegraphics[scale=.33]{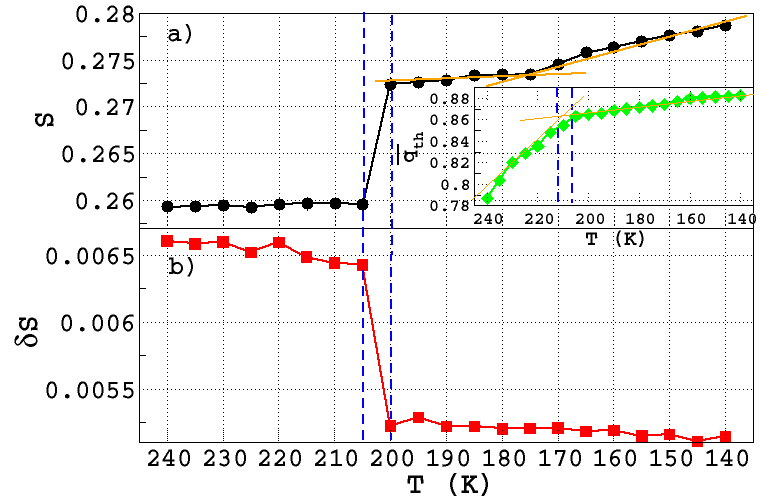}
    \includegraphics[scale=.33]{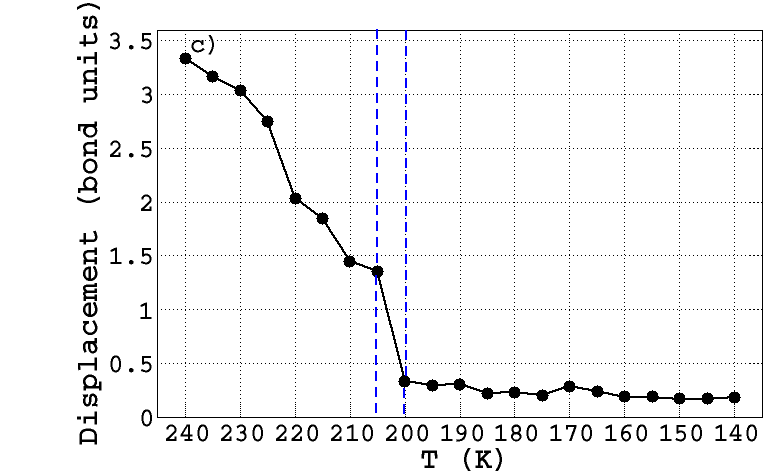}
    \caption{Evolution of $S$ (panel (a)) and $\delta S$ (panel (b)) during the water cooling protocol (see text). Panel (c) 
    reports the corresponding evolution of the standard molecular displacement in a ns time in units of the bond length. 
    The dashed vertical lines delimit the glass transition temperature ($T_g$) of the simulation. At $T=T_g$ the 
    translational motions freeze.}
 \label{fig:glass}
\end{figure}
In Fig.~\ref{fig:glass} we report the evolution of $S$ (panel (a)) and $\delta S$ (panel (b)) along the 
cooling protocol. Both OPs show a sudden, albeit small, change in correspondence with the vertical dashed blue lines. 
The sudden increase of $S$ indicates a sudden increase of the local order relative to that of the supercooled liquid. 
At the same time the sudden drop of $\delta S$ indicates reduced fluctuations of the local order relative to the 
supercooled liquid. The sharp variation of $S$ and $\delta S$ is associated to freezing of the translational motions 
in the system. This is illustrated in panel (c) of the same figure in which we report the standard displacement, i.e. 
the square root of the mean square displacement, of the molecules in a ns time, measured in units of the bond length 
(the nearest neighbor distance between the oxygen sites). While for $T\geq 205$ K the standard displacement is 
greater than $1$, indicating that each molecule on average moves by more than one bond length on a ns timescale, 
for $T\leq 200$ K the standard displacement drops well below $1$, indicating translational localization of the molecules. 
The freezing of the translational degrees of freedom marks the onset of the glass transition. The corresponding 
transition temperature, $T_g$, is located, in our simulation, in the interval bounded by the two dashed vertical lines. 
It is quite remarkable that a phenomenon usually associated with dynamics (viz. the freezing of translational
diffusion) has a clear static counterpart, well captured by the two OPs based on the LOM. The static signature of 
the glass transition is also detected by the average tetrahedral order parameter $\bar{q}_{th}$ shown in the inset of Fig.~\ref{fig:glass},
but in this case the effect is weaker as the transition is only signaled by a change of slope in the temperature 
variation of $\bar{q}_{th}$. Given that $\bar{q}_{th}$ weighs the tetrahedral order of the first shell of neighbors while the 
LOM focuses on the second shell of neighbors, we conclude that the second shell of neighbors provides a more 
sensitive gauge of the local order. 

\begin{figure}[!]
 \centering
    \includegraphics[scale=.33]{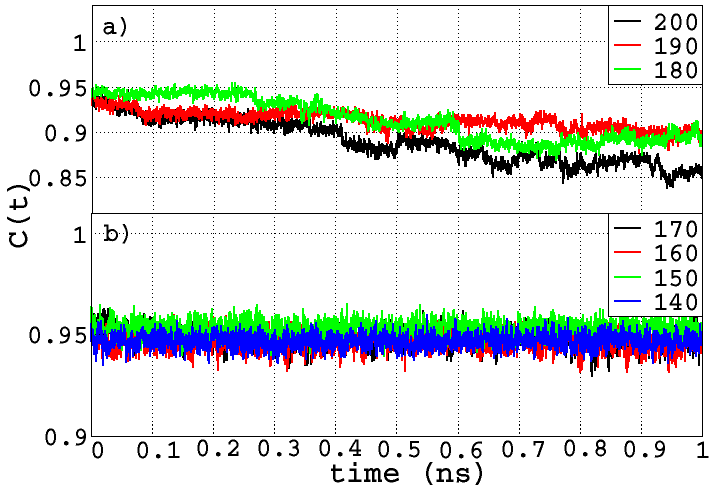}
    \caption{Time variation of the dipole autocorrelation function $C(t)$ in the temperature range $T\in[200, 180]$ K 
    (panel (a)) and in the temperature range $T\in[170, 140]$ K (panel (b)). Each $C(t)$ curve is obtained 
    by averaging 10 trajectories initiating at times separated by intervals of $100$ ps.}
 \label{fig:dipole}
\end{figure}

By further cooling the system below $T_g$ there is an evident change of slope in the increase of $S$ with temperature 
when $T$ is near $175$ K. No corresponding effect can be detected from the behavior of $\delta S$, which takes so 
small values to have lost sensitivity. 
This behavior is associated 
to the freezing of molecular rotations, as demonstrated in
Fig.~\ref{fig:dipole}, which reports the time evolution over $1$ ns of 
$C(t) = \left<\mathbf{\mu}(0)\cdot\mathbf{\mu}(t)\right>$, the time autocorrelation function of the molecular dipole
$\mathbf{\mu}$. The time decay of $C(t)$ is associated to rotational relaxation. 
In panel (a) of Fig.~\ref{fig:dipole} one sees that $C(t)$ decays on the ns timescale when $T\geq 180$ K, whereas 
for $T\leq 170$ K no apparent relaxation can be detected. We infer that freezing of the rotational degrees of 
freedom occurs when T is near $175$ K in our simulation. Similar rotational freezing effects have been inferred in recent 
experiments~\cite{shephard_molecular}.

\begin{figure}[!]
 \centering
    \includegraphics[scale=.33]{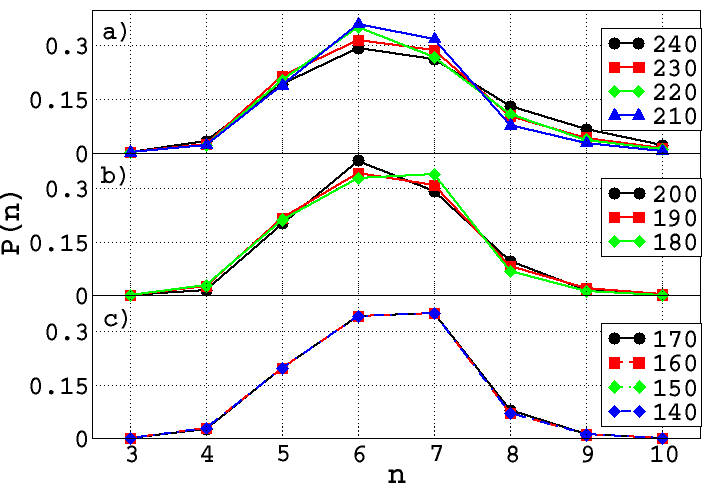}
    \caption{Ring distribution P(n) in the temperature range $T\in[240, 210]$ K (panel (a)), 
    in the temperature range $T\in[200, 180]$ K (panel (b)), 
    and in the temperature range $T\in[170, 140]$ K (panel (c)).}
 \label{fig:rings}
\end{figure}

To get further insight on the relaxation processes associated to the rotational and translational motions of the molecules, 
we analyzed the changes of the hydrogen bond network occurring upon cooling by monitoring the 
corresponding changes in the distribution $P(n)$ of the $n-$member rings in the network. Here $n$ indicates the number 
of hydrogen bonds in a ring. We define hydrogen bonds with the Luzar-Chandler 
criterion~\cite{luzar_hydrogen} and follow King's approach~\cite{king} for the ring statistics. We
report $P(n)$ at different temperatures in Fig.~\ref{fig:rings}. We see that $P(n)$ changes with temperature in the 
supercooled liquid (panel (a)) and also in the glass in the range of temperatures between $T_g$ and the 
temperature of rotational freezing (panel (b)), but below the latter no further changes in the topology of the 
network occur (panel (c)). As the temperature is lowered in the supercooled liquid longer member rings 
with $n\geq 8$ systematically disappear while the population of $6-$ and $7-$member
rings increases, with a prevalence of the former. Longer member rings are associated to more disordered local 
environments with interstitial molecules populating the region between the first and the second shell of 
neighbors~\cite{santra_local_2015}. Such configurations are typical of molecular environments with higher 
number density. As the temperature is lowered in the supercooled liquid at ambient pressure, this continuously transforms 
into a  liquid with lower density. In the glass, at temperatures above rotational freezing, network relaxation still occurs, 
again with a reduction in the population of rings with $n\geq 8$, but this time this is accompanied by a reduction 
of the population of the $6-$fold rings and an increase of the population of the $7-$fold rings. This is because 
the only processes that can change the network topology in absence of diffusion are bond switches of the 
kind described in Ref.~\cite{wooten_computer}. In water these processes can be generated by rotations
of the molecules. For instance, we found that in a frequent process of this kind 
two adjacent rings, a $6-$fold and an $8-$fold ring sharing a bond, transform into two adjacent $7-$fold rings sharing a bond. 
Finally, at temperatures below rotational freezing the network topology does not change in the timescale of the simulation. 
At these temperatures only local vibrational relaxation occurs. In a classical system vibrational disorder diminishes 
with temperature as reflected in the increase of $S$ at low temperature in Fig.~\ref{fig:glass} (a).   
\section{Conclusions}\label{conclusions}
We have introduced a local order metric (LOM) based on a simple \textit{measure} of the optimal overlap
between local configurations and reference patterns. In systems made of a repeated unit (atom or molecule) the LOM leads to the definition
of two global order parameters, $S$ and its spread $\delta S$, which have higher resolving power than popular alternative OPs and are
very useful to analyze structural changes in computer simulations, as shown by the 
examples in Section~\ref{results_1}, Section~\ref{Discerning} and Section~\ref{results_2}. 
The water examples show that the LOM can be used to measure the local order
not only at atomic but also at molecular sites. Systems made by molecular units
more complex than water could also be analyzed with this technique, while further
generalizations could be envisioned for binary and multinary systems.

As defined, $S$ and $\delta S$ are not
differentiable functions of the atomic (molecular) coordinates. This non-differentiability
stems from two reasons: (1) the $M$ neighbors of a site may change abruptly in a simulation, 
and (2) the LOM depends on the permutations of pattern indices, which is a
discrete variable. Thus $S$ and $\delta S$ could not be used as such to drive structural transformations
in constrained molecular dynamics simulations. However, they could be used as collective variables 
in Monte Carlo simulations adopting enhanced sampling
techniques, such as umbrella sampling~\cite{torrie_nonphysical_1977}, 
metadynamics~\cite{laio_escaping_2002}, replica exchange~\cite{swendsen_replica_1986} etc..
\\

\begin{acknowledgments}  
F.M., H-Y.K. and R.C. acknowledge support from the Department of Energy (DOE) under Grant No. DE-SC0008626.
E.C.O acknowledge financial support from the German Research Foundation (DFG) within the Postdoctoral Research 
Fellowship Program under Grant No. OG 98/1-1.
This research used resources of the National Energy Research Scientific Computing Center, which is supported by the 
Office of Science of the U.S. Department of Energy under Contract No. DE-AC02-05CH11231.
Additional computational resources were provided by the Terascale Infrastructure for Groundbreaking Research in 
Science and Engineering (TIGRESS) High Performance Computing Center and Visualization Laboratory at Princeton University.
\end{acknowledgments} 

\bibliographystyle{unsrt}
\linespread{0.1}
\bibliography{GeM_theory,GeM_additional}

\end{document}